\def\MPL #1 #2 #3 {Mod.~Phys.~Lett.~{\bf#1},\  #2 (#3)}
\def\NPB #1 #2 #3 {Nucl.~Phys.~{\bf#1},\  #2 (#3)}
\def\PLB #1 #2 #3 {Phys.~Lett.~{\bf#1},\  #2 (#3)}
\def\PR #1 #2 #3 {Phys.~Rep.~{\bf#1},\ #2 (#3)}
\def\PRD #1 #2 #3 {Phys.~Rev.~{\bf#1},\  #2 (#3)}
\def\PRL #1 #2 #3 {Phys.~Rev.~Lett.~{\bf#1},\  #2 (#3)}
\def\RMP #1 #2 #3 {Rev.~Mod.~Phys.~{\bf#1},\  #2 (#3)}
\def\ZP #1 #2 #3 {Z.~Phys.~{\bf#1},\  #2 (#3)}
\def\IJMP #1 #2 #3 {Int.~J.~Mod.~Phys.~{\bf#1},\  #2 (#3)}

\def\lam{\lambda}
\def\lamu{\lam_u}
\def\lamd{\lam_d}
\def\lamud{\lam_u^\dagger}
\def\lamdd{\lam_d^\dagger}
\def\mgut{M_U}

\def\mth{m_{3/2}}
\def\delgs{\delta_{GS}}
\def\kpr{K^\prime} 

\def\etc{{\it etc.}}
\def\sur{{\wt u_R}}
\def\msur{{m_{\sur}}}
\def\ibid{{\it ibid.}}
\def\anti{\overline}
\def\call{{\cal L}}

\def\del{\delta}
\def\half{{1\over 2}}

\def\rts{\sqrt s}

\def\lam{\lambda}

\def\eg{{\it e.g.}}

\def\epem{e^+e^-}
\def\mupmum{\mu^+\mu^-}

\def\lsim{\mathrel{\raise.3ex\hbox{$<$\kern-.75em\lower1ex\hbox{$\sim$}}}}
\def\gsim{\mathrel{\raise.3ex\hbox{$>$\kern-.75em\lower1ex\hbox{$\sim$}}}}
\def\@versim#1#2{\vcenter{\offinterlineskip
        \ialign{$\m@th#1\hfil##\hfil$\crcr#2\crcr\sim\crcr } }}

\def\slash#1{#1\hskip-6pt/\hskip2pt}

\def\etmiss{\slash E_T}

\def\ie{{\it i.e.}}

\def\gam{\gamma}

\def\anti{\overline}

\def\gev{\,{\rm GeV}}
\def\tev{\,{\rm TeV}}

\def\wt{\widetilde}

\def\mhalf{m_{1/2}}
\def\gl{\wt g}
\def\mgl{m_{\gl}}

\def\stopone{\wt t_1}
\def\stoptwo{\wt t_2}

\def\mstopone{m_{\stopone}}
\def\mstoptwo{m_{\stoptwo}}

\def\sbotone{\wt b_1}

\def\msbotone{m_{\sbotone}}

\def\stl{{\wt t_L}}
\def\str{{\wt t_R}}
\def\mstl{m_{\stl}}

\def\sbl{{\wt b_L}}
\def\sbr{{\wt b_R}}
\def\msbl{m_{\sbl}}
\def\msbr{m_{\sbr}}

\def\slepl{\wt \ell_L}
\def\mslepl{m_{\slepl}}
\def\slepr{\wt \ell_R}
\def\mslepr{m_{\slepr}}

\def\hl{h^0}

\def\mhl{m_{\hl}}

\def\tanb{\tan\beta}

\def\mt{m_t}

\def\mz{m_Z}

\def\mgut{M_U}

\def\cnone{\wt\chi^0_1}
\def\cntwo{\wt\chi^0_2}
\def\cnthree{\wt\chi^0_3}

\def\mcnone{m_{\cnone}}
\def\mcntwo{m_{\cntwo}}

\def\cpone{\wt \chi^+_1}
\def\cmone{\wt \chi^-_1}
\def\cpmone{\wt \chi^{\pm}_1}

\def\mcpmone{m_{\cpmone}}

\def\cpmtwo{\wt \chi^{\pm}_2}

\def\mcpmtwo{m_{\cpmtwo}}

\documentclass[smus]{snow2e}
\input psfig.sty

\begin{document}

\title{
{\normalsize UCD-96-25 \hspace*{\fill} September, 1996}\\
Motivations for and Implications of Non-Universal
GUT-Scale Boundary Conditions for Soft SUSY-Breaking 
Parameters.\thanks{
To appear in ``Proceedings of the 1996 DPF/DPB Summer Study
on New Directions for High Energy Physics''.
Work supported in part by the Department of Energy
and by the Davis Institute for High Energy Physics.
}\vskip -.1in}
\author{
G. Anderson (FNAL), C.H. Chen (U.C. Davis), 
J.F. Gunion (U.C. Davis), \\
J. Lykken (FNAL), T. Moroi (LBL), Y. Yamada (Wisconsin) 
\\ 
}

\maketitle

\thispagestyle{empty}\pagestyle{empty}

\begin{abstract} 
We outline several well-motivated models in
which GUT boundary conditions for SUSY breaking
are non-universal.  The diverse phenomenological implications
of the non-universality for SUSY discovery at LEP2,
the Tevatron, the LHC and the NLC are sketched.
\end{abstract}

\section{Introduction}

We will consider models in which the gaugino masses and/or the
scalar masses are not universal at the GUT scale, $\mgut$. The important
issues are: a) the extent to non-universal boundary conditions
can influence experimental signatures for the production
of supersymmetric particles and possibly suggest special
detector requirements needed to guarantee that the largest
possible class of supersymmetric models lead to observable
signatures at present and future lepton and hadron colliders; 
and b) the degree to which experimental data can distinguish
between different hypotheses/models for the
GUT scale boundary conditions. In this brief report, we attempt
to develop some insight into the answers to both questions
by focusing on some particularly well-motivated, but very
different, scenarios for the GUT scale boundary conditions.
At least one motivation for this report is to 
emphasize the fact that experimentalists
must not rely on the phenomenology of any one model in planning
their experiment or analyzing present or future data.  The alternative
possibilities presented here turn out to be relatively extreme in
some respects, and thus may provide useful benchmarks.  
However, we will be conservative in that we do not allow R-parity violation;
the LSP will always be the lightest neutralino and it will be invisible.

\section{Non-Universal Gaugino Masses at $\mgut$}

We focus on two different types of models 
in which gaugino masses are naturally not universal at $\mgut$.
\begin{itemize}
\item Superstring-motivated models in which SUSY breaking is
moduli (as opposed to dilaton) dominated.  We consider
the particularly attractive O-II model of Ref.~\cite{ibanez}
in which all matter fields are placed in the untwisted sector
and the universal `size' modulus field is the only source of SUSY
breaking. In this model, gaugino masses derive from one-loop
terms of a form that would be present in any theory. The boundary
conditions at $\mgut$ are:
\begin{equation}
\begin{array}{l}
M_a^0\sim \sqrt 3 \mth[-(b_a+\delgs)K\eta] \\
m_0^2=\mth^2[-\delgs\kpr] \\
A_0=0
\end{array}
\label{bcs}
\end{equation}
where $b_a$ are the standard gauge coupling RGE equation coefficients
($b_3=3$, $b_2=-1$, $b_1=-33/5$), $\delgs$
is the Green-Schwarz mixing parameter,
which would be a negative integer in the O-II model, with
$\delgs=-4,-5$ preferred; and $\eta=\pm1$.
The estimates of Ref.~\cite{ibanez} are
$K=4.6\times 10^{-4}$ and $\kpr=10^{-3}$, which
imply that slepton and squark masses would be very much
larger than gaugino masses.  It can be argued that the general
relation of the $M_a$ to $b_a$ and $\delgs$ is much more general
than the O-II model, and very likely to even survive
the non-perturbative corrections that will almost certainly
be present, whereas the relation between the $M_a$ constant $K$
and the $m_0$ constant $\kpr$ is much more model-dependent.
\item
Models in which SUSY breaking occurs via an $F$-term that is not
an SU(5) singlet.  In this class of models, gaugino masses are generated
by a chiral superfield $\Phi$ that appears linearly in the gauge
kinetic function, and whose auxiliary $F$ component acquires an
intermediate scale vev:
\begin{equation}
\call= \int d^2\theta W_a W_b {\Phi_{ab}\over M_{\rm Planck}} + h.c.
\sim  {\langle F_{\Phi} \rangle_{ab}\over M_{\rm Planck}}\lam_a\lam_b\,,
\end{equation}
where the $\lam_{a,b}$ ($a,b=1,2,3$) are the gaugino fields.  
If $F$ is an SU(5)
singlet, then $\langle F_{\Phi} \rangle_{ab}\propto c\del_{ab}$
and gaugino masses are universal.  
More generally, $\Phi$ and $F_{\Phi}$
need only belong to an SU(5) irreducible representation which
appears in the symmetric product of two adjoints:
\begin{equation}
({\bf 24}{\bf \times} 
{\bf 24})_{\rm symmetric}={\bf 1}\oplus {\bf 24} \oplus {\bf 75}
 \oplus {\bf 200}\,,
\label{irrreps}
\end{equation}
where only $\bf 1$ yields universal masses.  For the other representations
only the component of $F_{\Phi}$ that is `neutral' with respect to
the SM SU(3), SU(2) and U(1) groups should acquire a vev, assuming
that these groups remain unbroken after SUSY breaking. In this case
$\langle F_{\Phi} \rangle_{ab}=c_a\del_{ab}$, with $c_a$ depending
upon which of the above representations $\Phi$ lies in. 
The $c_a$ then determine the relative magnitude of 
the gauginos masses at $\mgut$. The results for the four possible
irreducible representations, Eq.~(\ref{irrreps}),
appear in Table~\ref{masses}. An arbitrary superposition
of the four irreducible representations is also, in principle,
possible.  In what follows, we shall assume that 
only one of the irreducible representations is present.
\end{itemize}

\begin{table}
\begin{center}
\begin{small}
\begin{tabular}{|c|ccc|ccc|}
\hline
\ & \multicolumn{3}{c|} {$\mgut$} & \multicolumn{3}{c|}{$\mz$} \cr
$F_{\Phi}$ 
& $M_3$ & $M_2$ & $M_1$ 
& $M_3$ & $M_2$ & $M_1$ \cr
\hline 
${\bf 1}$   & $1$ &$\;\; 1$  &$\;\;1$   & $\sim \;6$ & $\sim \;\;2$ & 
$\sim \;\;1$ \cr
${\bf 24}$  & $2$ &$-3$      & $-1$  & $\sim 12$ & $\sim -6$ & 
$\sim -1$ \cr
${\bf 75}$  & $1$ & $\;\;3$  &$-5$      & $\sim \;6$ & $\sim \;\;6$ & 
$\sim -5$ \cr
${\bf 200}$ & $1$ & $\;\; 2$ & $\;10$   & $\sim \;6$ & $\sim \;\;4$ & 
$\sim \;10$ \cr
\hline
 $\stackrel{\textstyle O-II}{\delgs=-4}$ & $1$ & $\;\;5$ & ${53\over 5}$ & 
$\sim 6$ & $\sim 10$ & $\sim {53\over5}$ \cr
\hline
\end{tabular}
\end{small}
\caption{Relative gaugino masses at $\mgut$ and $\mz$
for the four possible $F_{\Phi}$ irreducible representations
and in the O-II model with $\delgs\sim -4$.}
\label{masses}
\end{center}
\end{table}

\begin{figure}[htb]
\leavevmode
\begin{center}
\centerline{\psfig{file=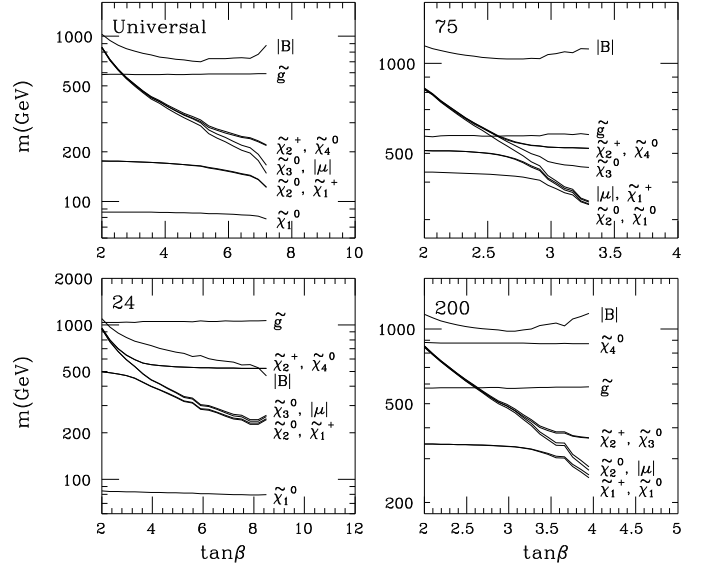,width=3.5in}}
\end{center}
\caption{Physical (pole) gaugino masses as a function of $\tanb$
for the ${\bf 1}$ (universal), ${\bf 24}$, ${\bf 75}$, and ${\bf 200}$
$F$ representation choices. Also plotted are $|B|$ and $|\mu|$.
We have taken $m_0=1\tev$ and $M_a(\mgut)=\mhalf$ times the
number appearing in Table~\ref{masses}, with $\mhalf=200\gev$.}
\label{mtanb}
\end{figure}

\begin{figure}[htb]
\leavevmode
\begin{center}
\centerline{\psfig{file=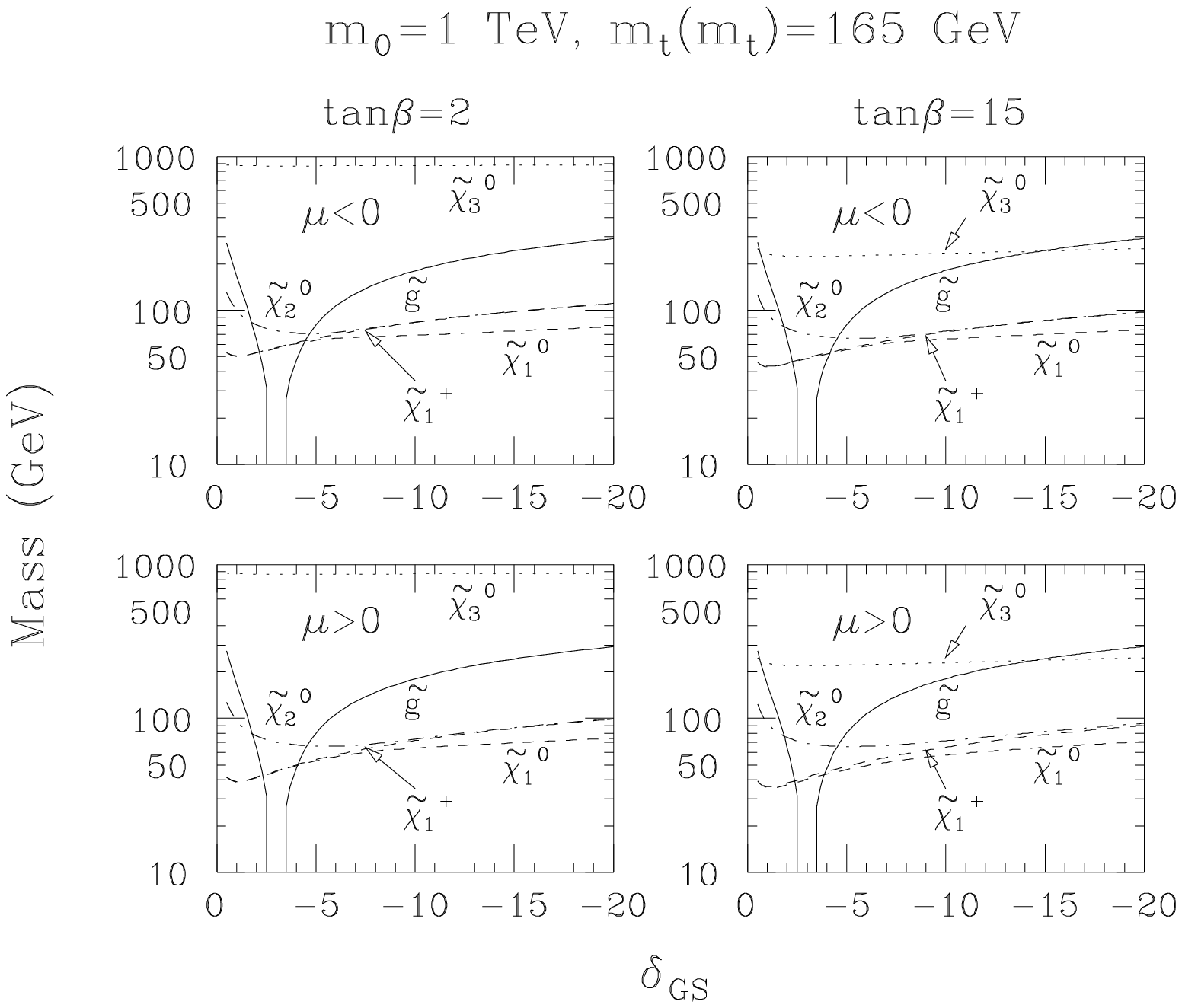,width=3.5in}}
\bigskip
\caption{Masses for the $\cnone$, $\cntwo$, $\cnthree$, $\cpone$,
and $\gl$ are plotted as a function of $\delgs$ at $\tanb=2$ and 15
and for $\mu>0$ and $\mu<0$.
}
\label{inos}
\end{center}
\end{figure}

To understand the implications of GUT-scale choices for the $M_a$, we
need only recall that $M_3^0:M_2^0:M_1^0$ at $\mgut$ evolves
to $M_3:M_2:M_1=3M_3^0:M_2^0:\half M_1^0$ at $\mz$, as indicated
in Table~\ref{masses}. Physical masses of the gauginos are
influenced by $\tanb$-dependent off diagonal terms in the mass
matrices and by corrections which boost $\mgl(pole)$ relative to $\mgl(\mgl)$.
If $\mu$ is large, the lightest neutralino (which is the LSP)
will have mass $\mcnone\sim {\rm min}(M_1,M_2)$ while the lightest
chargino will have $\mcpmone\sim M_2$. Thus, in the ${\bf 200}$ 
and O-II scenarios with
$M_2\lsim  M_1$, $\mcpmone\simeq\mcnone$ and the $\cpmone$
and $\cnone$ are both wino-like. The
$\tanb$ dependence of the masses at $\mz$ for the universal,
${\bf 24}$, ${\bf 75}$, and ${\bf 200}$ choices appears in Fig.~\ref{mtanb}.
The gaugino masses in the O-II scenario are plotted
as a function of $\delgs$ for $\tanb=2$ and 15 in Fig.~\ref{inos}.
The $\mgl-\mcnone$ mass splitting becomes increasingly
smaller in the sequence ${\bf 24}$, ${\bf 1}$, ${\bf 200}$
${\bf 75}$, O-II ($\delgs\sim -4$), 
as could be anticipated from Table~\ref{masses}.
Indeed, in the O-II case $\mgl< \mcnone$ (even after including the gluino 
pole mass correction) at $\delgs=-4$; $\delgs\gsim-4.2$ yields $\mgl>\mcnone$
by just a small amount.
It is interesting to note that at
high $\tanb$ values $\mu$ decreases to a level comparable to $M_1$ and
$M_2$, and there is substantial degeneracy among the $\cpmone$, $\cntwo$
and $\cnone$.

The two models described above are certainly not the only ones
in which gaugino masses are non-universal. For example,
in Ref.~\cite{doublettriplet} a scenario is developed 
in which non-universal gaugino masses
are intimately related to a solution of the doublet-triplet
Higgs mass splitting naturalness problem \cite{HIY}.  One finds:
\begin{equation}
{M_{a}\over g_{a}^2}= {M_{H_{a}}\over k_a g_{H_{a}}^2}+{M_G\over g_G^2}\,,~~~
(a=1,2,3)\,,
\end{equation}
where $M_{H_2}=0$, $k_3=1$, $k_1=15$, 
$G$ is the usual SU(5) group, $H_1$ and $H_3$ 
are new groups with associated coupling constants $g_{H_{1,3}}$,
the $g_a$ are the usual coupling constants, 
and all $M/g^2$ ratios are RG invariants. Since there are essentially no
constraints on $M_{H_{1,3}}/g_{H_{1,3}}^2$ at $\mgut$, the gaugino masses
become independent parameters.

\section{Non-universal scalar masses at $\mgut$}

Two classes of model come immediately to mind.
\begin{itemize}
\item
Models in which the SUSY-breaking scalar masses at $\mgut$ are influenced
by the Yukawa couplings of the corresponding quarks/leptons.  
This idea is exemplified in the model of Ref.~\cite{hallrandall}
based on perturbing about the $[U(3)]^5$ symmetry that is present
in the absence of Yukawa couplings.  One finds:
\begin{eqnarray}
&m_{\wt Q}^2=m_0^2(I+c_Q\lamud\lamu+c_Q^\prime\lamdd\lamd+\ldots) &\\
&m_{\wt U^c}^2=m_0^2(I+c_U\lamu\lamud+\ldots) &\\
&m_{\wt D^c}^2=m_0^2(I+c_D\lamd\lamdd+\ldots) &
\end{eqnarray}
where $Q$ represents the squark partners of the left-handed quark doublets
and $U^c~(D^c)$ the partners of the left-handed up (down) antiquarks.
The $m^2$'s, $I$  and $\lamu,\lamd$ are $3\times 3$ matrices
in generation space, the latter containing the Yukawa couplings of the 
quarks. The $\ldots$ represent terms of order $\lam^4$ that we will neglect.
A, $c_Q$, $c_Q^\prime$, $c_U$ and $c_D$ should all be similar in size,
in which case the large top quark Yukawa coupling implies that
the primary deviations from universality will occur in $\mstl^2$,
$\msbl^2$ (equally and in the same direction) and $\msbr^2$.
We also recall that $A$ terms can be present that will mix the $Q$
and $U^c,D^c$ squark fields, with
\begin{equation}
A_U=A_0(\lamu+\ldots)\,, \quad A_D=A_0(\lamd+\ldots)\,,
\end{equation}
where $A_U$ ($A_D$) describes $Q-U^c$ ($Q-D^c$) mixing, respectively, and
$\ldots$ represents cubic and higher terms that we neglect.
It is the fact that $\mstl^2$ and $\msbl^2$ are equally shifted
that will distinguish $m^2$ non-universality from the effects of a large
$A_0$ parameter at $\mgut$; the latter would primarily 
introduce $\stl-\str$ mixing and yield a low
$\mstopone$ compared to $\msbotone$.
\item
A second source of non-universal scalar masses is closely related
to the second non-universal gaugino mass scenario discussed earlier.
In close analogy, scalar masses will arise from the effective
Lagrangian form:
\begin{equation}
\call\propto {\langle F_{\Phi} F_{\Phi}\rangle_{ij}\over M_{\rm Planck}^2}
\phi_i^\dagger\phi_j\,,
\end{equation}
where the $\phi_{i,j}$ are the scalar fields
and we implicitly assume that only a single $F_{\Phi}$ 
in an irreducible representation of SU(5) is active.
Since the $\phi_i$'s associated with the partners of
the (left-handed) SM fermion and antifermion fields
appear in both $\overline{\bf 5}$ and $\bf 10$ representations, 
while the doublet Higgs fields that must remain light appear in a $\bf 5$,
the representation $\bf R$ of $\Phi$ must be chosen so that
${\bf R\times\bf R}$ overlaps one or more of the representations common to
$\bf 10\times \overline{\bf 10}={\bf 1}\oplus{\bf 24}\oplus{\bf 75}$ and 
$\bf 5\times \overline{\bf 5}={\bf 1}\oplus{\bf 24}$. For example, 
${\bf R}={\bf 1}$ and [see Eq.~(\ref{irrreps})] ${\bf R}={\bf 24}$
both would work, and illustrate the possibility
that a single $F_{\Phi}$ could simultaneously give rise to both gaugino
and squark soft mass terms. The analysis of squark masses
in the general situation is complex and will be left to a future work.  
There are clearly many possible patterns of non-universality.
\end{itemize}
Finally, we note that universality is predicted for the scalar masses
in the O-II model, although in modest variants thereof a
limited amount of non-universality can be introduced.

\section{Phenomenology}

We separately
consider gaugino mass non-universality and squark mass non-universality,
although the two could be
interrelated in the $F_{\Phi}$ models. Even if these
different types of non-universality do not derive from the same mechanism,
both could be simultaneously present. We note that we found it to be very
straightforward to incorporate alternate non-universal 
boundary conditions into event generators such as ISASUGRA/ISASUSY. 
Interested parties are encouraged to contact us for specific instructions.

\subsection{Non-universal gaugino masses}

The gaugino mass patterns outlined in Table~\ref{masses}
have important phenomenological
implications, only a few of which we attempt to sketch here.
\begin{itemize}
\item For $\mcpmone\sim \mcnone$ (${\bf 200}$, O-II), in $\cpone\to\cnone
\ell\nu,\cnone 2j$ decays the $\ell$ and jets are very soft implying:
\begin{enumerate}
\item
$\epem \to \cpone\cmone$ detection may require using a 
photon tag \cite{invisible};
\item the like-sign signal for $\gl\gl$ production disappears;
\item the tri-lepton signal for $\cpmone\cntwo$ disappears.
\end{enumerate}
\item For $\mgl\sim\mcpmone\sim\mcnone$ (O-II, $\delgs\sim -4$), 
the $jets+\etmiss$ signal
for $\gl\gl$ production is more difficult to extract --- softer
jet cuts must be employed than in other scenarios, 
implying larger backgrounds, see \cite{moddominated};
\item Decay patterns and mass ratios are a strong function
of scenario, implying experiment can distinguish different
scenarios from one another.
\end{itemize}

It is particularly amusing to examine the phenomenological 
implications of these
boundary conditions for the standard Snowmass overlap point specified
by $\mt=175\gev$, $\alpha_s=0.12$,
$m_0=200\gev$, $M_3^0=100\gev$\footnote{We fix $M_3$
to be the same in all scenarios so that $\mgl$ will
have roughly the same value in all models.}
$\tanb=2$, $A_0=0$ and $\mu<0$.
For the given $M_3^0$, $m_0$ is very large in the O-II scenario
if the strict 1-loop values of $K$ and $\kpr$ noted earlier are employed. 
However, the relation between $K$ and $\kpr$ could be drastically altered
by non-perturbative corrections.  Thus, in treating the O-II model
below, we shall take $m_0=600\gev$, a value that yields a (pole)
value of $\mgl$ not unlike that for the other scenarios; $m_0=200\gev$
would imply that the $\cnone$ would not be the LSP.
By comparing these scenarios we can gain a first insight as
to the degree to which experiment will allow us to determine
the appropriate GUT/String scale boundary conditions.

The masses of the supersymmetric particles for
each scenario are given in Table~\ref{susymasses}.
As promised, the $\cpmone$ is very degenerate with
the $\cnone$ in the $\bf 200$ and O-II models. 

\begin{table}
\begin{center}
\begin{small}
\begin{tabular}{|c|c|c|c|c|c|}
\hline
\ & ${\bf 1}$ & ${\bf 24}$ & ${\bf 75}$ & ${\bf 200}$ & 
$\stackrel{\textstyle O-II}{\delgs=-4.7}$ \cr
\hline 
$\mgl$          & 285 & 285 & 287 & 288 & 313 \cr
$\msur$         & 302 & 301 & 326 & 394 &  -   \cr
$\mstopone$     & 255 & 257 & 235 & 292 &   -   \cr
$\mstoptwo$     & 315 & 321 & 351 & 325 &   -   \cr
$\msbl$         & 266 & 276 & 307 & 264 &   -   \cr
$\msbr$         & 303 & 303 & 309 & 328 &   -   \cr
$\mslepr$       & 207 & 204 & 280 & 437 &   -   \cr
$\mslepl$       & 216  & 229 & 305 & 313 &   -   \cr
$\mcnone$       & 44.5 & 12.2 & 189 & 174.17  &   303.09 \cr
$\mcntwo$       & 97.0 & 93.6 & 235 & 298 &   337 \cr
$\mcpmone$      & 96.4 & 90.0 & 240 & 174.57 &   303.33 \cr
$\mcpmtwo$      & 275 & 283  & 291 & 311 &    - \cr
$\mhl$          & 67  & 67   & 68   & 70   &   82 \cr
\hline
\end{tabular}
\end{small}
\caption{Sparticle masses for the Snowmass comparison point
in the different gaugino mass scenarios. Blank entries for the O-II
model indicate very large masses.}
\label{susymasses}
\end{center}
\end{table}

The phenomenology of these scenarios for $\epem$ collisions
is not absolutely straightforward. 
\begin{itemize}
\item
In the $\bf 1$ and $\bf 24$ models the
masses for the $\slepl$, $\slepr$, $\cpmone$, $\cnone$ and $\cntwo$ are
such that the standard array of SUSY discovery channels at the NLC
would be present and easily observable since all mass splittings
are substantial.
\item
In the $\bf 75$ model, $\cpone\cmone$ and $\cntwo\cntwo$
pair production at $\rts=500\gev$ are
barely allowed kinematically; the phase space for $\cnone\cntwo$
is only somewhat better.  All the signals would be rather weak,
but could probably be extracted with sufficient integrated luminosity.
It might prove fruitful to look for $\epem\to\gam\cnone\cnone$.
\item
In the $\bf 200$ model, $\epem\to \cpone\cmone$ production would be 
kinematically allowed at a $\rts=500\gev$ NLC, but not easily observed
due to the fact that the (invisible) $\cnone$ would take essentially all of
the energy in the $\cpmone$ decays.  However, 
according to the results of Ref.~\cite{invisible}, 
$\epem\to \gam\cpone\cmone$ would be observable at $\rts=500\gev$. 
The only other directly visible (\ie\ without a $\gam$ tag)
sparticle pair channel would be $\epem\to \cntwo\cnone$; the small
phase space would imply a very weak signal.
\item
The O-II model with $\delgs$ near $-4$ predicts that $\mcpmone$ 
and $\mcnone$ are both rather close to $\mgl$, so that
$\epem\to \cpone\cmone,\cnone\cnone$ would {\it not} be kinematically allowed.
The only SUSY `signal' would be the presence of a very SM-like
light Higgs boson.
\end{itemize}

At the LHC, the strongest signal for SUSY would arise from $\gl\gl$
production. The different models lead to very distinct signatures
for such events.  To see this, it is sufficient to list the primary
easily identifiable decay chains of the gluino for each of the five scenarios.
(In what follows, $q$ denotes any quark other than a $b$.)
\begin{eqnarray*}
 {\bf 1:} && \gl \stackrel{90\%}{\to} \sbl\anti b \stackrel{99\%}{\to}
\cntwo b\anti b\stackrel{33\%}{\to} \cnone(\epem~{\rm or}~\mupmum)b\anti b\\
 \phantom{{\bf 1:}} && 
\phantom{\gl \stackrel{90\%}{\to} \sbl\anti b \stackrel{99\%}{\to}
\cntwo b\anti b}
\stackrel{8\%}{\to} \cnone \nu\anti\nu b\anti b\\
 \phantom{{\bf 1:}} && 
\phantom{\gl \stackrel{90\%}{\to} \sbl\anti b \stackrel{99\%}{\to}
\cntwo b\anti b}
\stackrel{38\%}{\to} \cnone q\anti q b\anti b\\
 \phantom{{\bf 1:}} && 
\phantom{\gl \stackrel{90\%}{\to} \sbl\anti b \stackrel{99\%}{\to}
\cntwo b\anti b}
\stackrel{8\%}{\to} \cnone b\anti b  b\anti b\\
 {\bf 24:} && \gl \stackrel{85\%}{\to} \sbl\anti b
\stackrel{70\%}{\to} \cntwo b\anti b\stackrel{99\%}{\to}
 \hl \cnone b\anti b\stackrel{28\%}{\to} \cnone b\anti b b\anti b \\
\phantom{ {\bf 24:}} && 
\phantom{ \gl \stackrel{85\%}{\to} \sbl\anti b
\stackrel{70\%}{\to} \cntwo b\anti b\stackrel{99\%}{\to}
 \hl \cnone b\anti b}
\stackrel{69\%}{\to} \cnone \cnone\cnone b\anti b \\
 {\bf 75:} && \gl \stackrel{43\%}{\to} \cnone g~{\rm or}~\cnone q\anti q \\
\phantom{{\bf 75:}} && \phantom{\gl}\stackrel{10\%}{\to} \cnone b\anti b \\
\phantom{{\bf 75:}} && \phantom{\gl}\stackrel{20\%}{\to} 
\cntwo g~{\rm or}~\cntwo q\anti q \\
\phantom{{\bf 75:}} && \phantom{\gl}\stackrel{10\%}{\to} \cntwo b\anti b \\
\phantom{{\bf 75:}} && \phantom{\gl}\stackrel{17\%}{\to} \cpmone q\anti q \\
 {\bf 200:} && \gl \stackrel{99\%}{\to} \sbl\anti b
\stackrel{100\%}{\to} \cnone b\anti b \\
 {\rm O-II{\bf:}} && \gl \stackrel{51\%}{\to} \cpmone q\anti q \\
\phantom{{\rm O-II{\bf:}}} && \phantom{\gl} \stackrel{17\%}{\to} \cnone g \\ 
\phantom{{\rm O-II{\bf:}}} && 
\phantom{\gl} \stackrel{26\%}{\to} \cnone q\anti q \\ 
\phantom{{\rm O-II{\bf:}}} && 
\phantom{\gl} \stackrel{6\%}{\to} \cnone b\anti b
\end{eqnarray*}

Gluino pair production will then lead to the following strikingly
different signals.
\begin{itemize}
\item In the $\bf 1$ scenario we expect a very large
number of final states with missing energy,
four $b$-jets and two lepton-antilepton pairs.
\item For $\bf 24$, an even larger number of events will have missing energy
and eight $b$-jets, four of which reconstruct to two pairs with
mass equal to (the known) $\mhl$.
\item
The signal for $\gl\gl$ production in the case of $\bf 75$ is
much more traditional; the primary decays yield
multiple jets (some of which are
$b$-jets) plus $\cnone$, $\cntwo$ or $\cpmone$. 
Additional jets, leptons and/or neutrinos
arise when $\cntwo\to\cnone$ + two jets,
two leptons or two neutrinos or $\cpmone\to\cnone$ +
two jets or lepton+neutrino.
\item
In the $\bf 200$ scenario, we find missing energy plus four $b$-jets; 
only $b$-jets appear in the primary decay --- any other
jets present would have to come from initial or final state radiation,
and would be expected to be softer on average. This is almost
as distinctive a signal as the $8b$ final state found in the $\bf 24$
scenario.
\item
In the final O-II scenario, $\cpmone\to \cnone$ + very soft
spectator jets or leptons that would not be easily detected.  Even
the $q\anti q$ or $g$ from the primary decay would not be very energetic
given the small mass splitting between $\mgl$ and $\mcpmone\sim\mcnone$.
Any energetic jets would have to come from initial or final state
radiation.  Soft jet cuts would have to be used to dig out this signal,
but it should be possible given the very high $\gl\gl$ production rate
expected for this low $\mgl$ value; see Ref.~\cite{moddominated}.
\end{itemize}
Thus, for the particular $m_0$,  $M_3$, $\tanb$, \etc\ values chosen for
the Snowmass comparison point, 
distinguishing between the different boundary condition scenarios 
at the LHC will be extremely easy.  Further, the event rate
for a gluino mass this low is such that the end-points of
the various lepton, jet or $\hl$ spectra will allow relatively good
determinations of the mass differences between the sparticles appearing
at various points in the final state decay chain \cite{endpoints}.
We are optimistic that this will prove to be 
a general result so long as event rates are large.

\subsection{Non-universal scalar masses}

In this section, we maintain
gaugino mass universality at $\mgut$, but allow for non-universality
for the squark masses, in which case it is natural to
focus on LHC phenomenology.

\begin{figure}[htb]
\leavevmode
\begin{center}
\centerline{\psfig{file=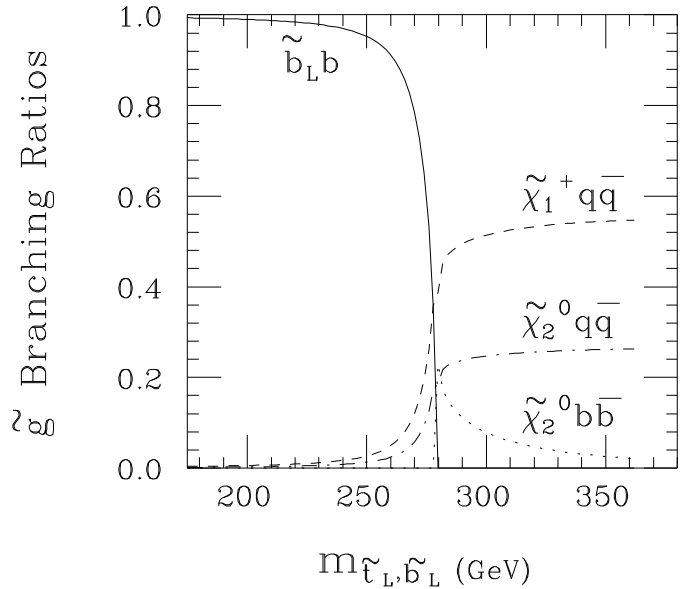,width=3.5in}}
\bigskip
\caption{Gluino branching ratios as a function
of $\mstl$ in the case of $c_Q\neq 0,c_{U}=c_D=A_0=0$. 
We employ the Snowmass overlap point
values of $M_a^0=100\gev$ ($a=1,2,3$), $m_0=200\gev$, $\tanb=2$, with $\mu<0$.
}
\label{mtlbr}
\end{center}
\end{figure}

We perturb about the Snowmass overlap point,
taking $c_Q\neq 0$ with $c_U=c_D=A_0=0$.
In Fig.~\ref{mtlbr} we plot the $\gl$ branching ratios as a function
of $\mstl=\msbl$ as $c_Q$ is varied from negative to positive values.
As the common mass crosses the threshold above which
the $\gl\to \sbotone  b$ decay becomes kinematically disallowed,
we revert to a more standard SUSY scenario in which $\gl$
decays are dominated by modes such as $\cpmone q\anti q$, 
$\cnone q\anti q$ (not plotted),
$\cntwo q\anti q$ and (more exotically) $\cntwo b\anti b$,
not unlike the $\bf 75$ non-universal gaugino mass
scenario as far as the important channels are concerned. To distinguish
between such a $c_Q\neq 0$ case vs. the $\bf 75$
scenario would require using the lepton/jet spectra in the final state
to determine if the $\cnone$ and $\cpmone$ are light vs. heavy,
respectively. (Of course, at the NLC the light $\cpmone$ present
in the $c_Q\neq 0$, universal-gaugino-mass scenario would be immediately
detected and its mass easily measured.)
As $\msbotone$ and $\mstopone$ increase in the $c_Q\neq 0$ scenario, 
the $\sbotone$ and $\stopone$ branching ratios also change; \eg\
$\sbotone\to \gl b$ goes from zero to 
dominant as the $\sbotone\to \cntwo b,\cmone t$ modes decline.

Experimental determination
of the squark masses will be very important for deciding if
corrections to scalar mass universality are present at $\mgut$
and for illuminating their nature.

\section{Conclusions}

By combining well-motivated 
GUT-scale gaugino mass non-universality
and squark mass non-universality scenarios an enormous array of
boundary conditions at $\mgut$ becomes possible.
Thus, it will be dangerous to focus on one signal/channel or
even short list of channels for discovery of supersymmetry.
Indeed, a thorough search and determination
of the rates (or lack thereof) for the full panoply of possible
channels is required to distinguish the many possible 
GUT-scale boundary conditions from one another. 

%


\begin{thebibliography}{2}

\bibitem{ibanez} A. Brignole, L.E. Ibanez, and C. Munoz, \NPB B422 125 1994 ,
Erratum, \ibid, {\bf B436} 747 (1995).
A. Brignole, L.E. Ibanez, C. Munoz, and C. Scheich,
hep-ph/9508258.

\bibitem{invisible} C.-H. Chen, M. Drees, and J.F. Gunion,
\PRL 76 2002 1996 .

\bibitem{moddominated} C.-H. Chen, M. Drees, and J.F. Gunion, 
hep-ph/9607421.

\bibitem{doublettriplet}  N. Arkani-Hamed, H.-C. Cheng, and T. Moroi,
hep-ph/9607463.

\bibitem{HIY} 	T. Hotta, K.-I. Izawa and T. Yanagida, 	\PRD D53 3913 1996 .

\bibitem{hallrandall} L.J. Hall and L. Randall, \NPB B352 289 1991 .


\bibitem{endpoints} 
R.M. Barnett, J.F. Gunion and H.E. Haber, \PLB B315 349 1993 ;
see also the supersymmetry group reports in this volume.

\end{thebibliography}
\end{document}